\documentstyle[12pt]{article}
\begin{document}
\title{Excitation of plasma oscillations by moving Josephson vortices in
layered superconductors}
\author{S.N. Artemenko and S.V. Remizov}
\maketitle
Institute for Radioengineering and Electronics of Russian Academy of
Sciences, 103907 Moscow, Russia
\footnote[1]{E-mail: Art@mail.cplire.ru}

\begin{abstract}
Electric and magnetic fields created by moving uniform lattice of Josephson
vortices in the magnetic field parallel to the layers are calculated in the
frame of exactly solvable model. At large velocities of the vortex lattice
the plasma oscillations of superconducting electrons are excited by vortex
motion, this results in interference features in I-V curves at low
temperatures. The spectrum of electromagnetic radiation by moving vortices
contains peaks related to the excitation of plasmons and to the Cherenkov
radiation.
\end{abstract}

It is known that in high-T$_c$ superconductors there exist low-damping
plasma oscillations \cite{v}-\cite{b} in the range of hundreds megahertz -
terahertz. The presence of an underdamped eigenmode leads to a resonance
behavior of the driven modes when frequency and wave-vector of the driven
force are close to those of the eigenmode. In particular, frequencies of the
electromagnetic field of sliding lattice of Josephson vortices created by
magnetic field parallel to the layers can enter the range of plasma modes if
the lattice is moving fast enough. The velocity of such vortices driven by
transport current perpendicular to the layers may be very large, because the
amplitude of the order parameter $\Delta$ inside Josephson vortices is
perturbed very slightly. Thus, at large enough voltages across the
superconductor the resonance conditions can be satisfied. This results in
the generation of plasma oscillations and influences the I-V curves. Below
we study the motion of uniform lattice of Josephson vortices in an infinite
crystal in magnetic field which is small enough, so that non-linear cores of
the vortices do not overlap. We assume that the characteristic frequencies
of the problem are small in comparison to $\Delta$.

In the presence of vortices the spectrum of the eigenmodes is different from
the zero-field spectrum due to oscillations of vortex displacements leading
to the sound-like \cite{v},\cite{b},\cite{s} spectrum of the modes. In
contrast, the response of a superconductor to the motion of the vortex
lattice driven by the transport current calculated in our study, is not
related to the oscillations of the vortices, driven modes being related to
the plasmons with the plasma edge.

The current distribution in the vortices is influenced by their motion, this
makes calculations of the vortex dynamics difficult since usual
perturbational approach is not valid at large velocities of the lattice. To
solve the problem strictly one must find solution of Maxwell's equations
with expressions for charge and current densities in layered
superconductors. Having in mind high-T$_c$ superconductors we shall consider
the order parameter having the symmetry close to d-pairing. Expressions for
current and charge densities are non-linear functions of the phase differences
between adjacent layers, which are rather complicated functions \cite{ij} in
general case. Quasi-particle current densities depend on phase differences,
on the supecronducting momentum ${\bf P}_n=(1/2)i{\bf q} \chi_n-(1/c){\bf
A}_n$ and, in addition, on the gauge-invariant scalar potential $\mu_n
=(1/2) (\partial \chi_n /\partial t)+\Phi_n$, where ${\bf A}_n$ is vector
potential, $\Phi_n$ is electric potential, and $\chi_n$ is the phase of the
order parameter in layer $n$. (We assume here $\hbar=1, e=1$). In general
case potential $\mu_n$ to be found from the Poisson's equation, however in
the vortex problem at low temperatures considered here, $T \ll \Delta$, the
quasi-particle branch imbalance, and, hence, potential $\mu_n$ can be
neglected provided $(r_0/d)^2 \ll 1$, where $r_0$ is Thomas-Fermi screening
radius, and $d$ is the period of the crystal in the direction perpendicular
to the layers. Current density along the layers can be described by
expression for linear response since the characteristic value of the current
is determined by the critical current density $j_c$ in the perpendicular
direction, which is small in comparison to the critical current in the
parallel direction. Then the expression for current density in layer $n$,
presented by means of Fourier transformation has the form $${\bf j}_n
=\frac{c^2}{4\pi \lambda^2}{\bf P}_n- i\omega\sigma_{\|}(\omega){\bf P}_n,$$
The first term here describes the superconducting current, and the second is
related to the quasi-particle current, $-i \omega$ and $i{\bf q}$ correspond
to the time derivative and to the gradient in the direction parallel to the
layers. The frequency dependence of conductivity $\sigma_{\|}(\omega)$ is
determined by symmetry of the order parameter and depends on the momentum
scattering time. We shall use the expressions from Ref.\cite{lr} to describe
$\sigma_{\|}(\omega)$.

The density of the superconducting current between layers $n$ and $n+1$ is
given by $$j^{(s)}_{\perp n}(\varphi_n)=j_c \sin{\varphi_n},$$
where $\varphi_n$ is the gauge invariant phase difference. We shall adopt an
exactly solvable model substituting $\sin{\varphi_n}$ by the saw-tooth
function
\begin{equation}
j^{(s)}_{\perp n}(\varphi_n)=j_c\arcsin{\sin{\varphi_n}} ,
\label{s}
\end{equation}
as it was done in \cite{Au}. Such a model was used to study the Josephson
vortex flow in the linear approximation on velocity in \cite{V}. Following
(\ref{s}) we shall use linear relation between the quasi-particle current and
the phase difference
$$j^{(qp)}_{\perp n}= -i\omega\sigma_{\perp}(\omega)\varphi_n/(2d).$$
The expressions for the current densities we insert to the Maxwell equation
\begin{equation}
{\rm rot}{\bf H}=\frac{4\pi}{c}{\bf j} + \frac{1}{c}
\frac{\partial {\bf D}}{\partial t},
\label{M}
\end{equation}
in which we shall use discrete approximation for spatial derivatives in $z$
direction perpendicular to the layers, and express magnetic field in $y$
direction as $\varphi_n$ and ${\bf P}_n$ $$H_y=\frac{c}{2d} \frac{\partial
\varphi_n}{\partial x} -\frac{c}{d}(P_{n+1}-P_n),$$ where $x$ is the 
coordinate in the direction of the vortex flow.

In the displacement current we take into account only the component along
$z$ axis, since due to the strong anisotropy of a layered superconductor the
plasma frequency in the direction parallel to the layers, $\Omega_p =c/
\lambda$, is much larger, than typical frequencies of the problem, which
are of the order of the plasma frequency for perpendicular direction,
$\omega_p= c_z/\lambda_c$. Here $c_z =c/\sqrt{\epsilon}$, $\epsilon$ is a
dielectric constant in $z$ direction, $\lambda$ and $\lambda_c =c/
\sqrt{8\pi d j_c}$ are the screening lengths for the current flowing
parallel and perpendicular to the layers, respectively. Then we get the
equations
\begin{eqnarray}
j^{(s)}_{\perp n}(\varphi_n)-\frac{c^2}{8\pi d}\frac{\partial^2
\varphi_n}{\partial x^2} +2\lambda_c^2\frac{\partial}{\partial x} (P_{n+1} -
P_n) =-\frac{\lambda_c^2}{c^2} \frac{\partial}{\partial t}
\left(\frac{4\pi\sigma_{\perp}}{\epsilon}+ \frac{\partial}{\partial t}
\right)\varphi_n
\label{1} \\
\frac{c^2}{2d^2}\frac{\partial}{\partial x}
(\varphi_n -\varphi_{n-1}) +\Omega_p^2 P_n -
\frac{c^2}{d^2}(P_{n+1}+P_{n-1}-2P_n)= -\frac{\partial}{\partial t}
\left(4\pi \sigma_{\|} +\frac{\partial}{\partial t} \right) P_n.
\label{2}
\end{eqnarray}
The solution of these equations yields spatial and temporal dependencies of
electric and magnetic fields.

Using the model current-phase relation (\ref{s}) we can find the exact
solution of equations (\ref{1}-\ref{2}) by means of the Fourier
transformation. In principle, we must take into account a bending of the
vortices due to their motion. However, we shall limit our study here by the
case when the displacement of the vortex center $x_0(y)$ along $x$ axis is
small at distances $y \approx \lambda_c$. Estimation of the vortex bending
based on equations of a balance between forces acting to the vortices
and created by the transport current and by the vortex currents shows that
the deformation of vortices can be neglected in magnetic fields $H \gg
H_{c1}$ provided
\begin{equation}
\frac{x_0(\lambda_c)}{\lambda_c} \approx \frac{\lambda}{\pi d
\ln{(\lambda/d)}} \frac{j_{tr} H_{c1}}{j_c H} \ll 1. \label{d}
\end{equation}
We assume that $H$ is large in comparison to $H_{c1}$ and condition
(\ref{d}) is satisfied.

Then for the triangle lattice sliding with velocity $u$ we get
\begin{eqnarray}
(\tilde{\omega}_p^2-\omega^2+i\omega\omega_r)\varphi-2c^2qKP/\epsilon=
\Pi\delta(q-\omega/u)
\label{3} \\
-(c^2qK/2d^2)\varphi
+(\tilde{\Omega}_p^2-\omega^2+i\omega\Omega_r)P =0.
\label{4}
\end{eqnarray}
Here $K=2\sin{k/2}$, $|k|<\pi$ is the wave number obtained from the discrete
Fourier transformation with respect to the layer number, $\tilde{\omega}_p^2=
\omega_p^2 (1+\lambda_c^2q^2)$, $\tilde{\Omega}_p^2=\Omega_p^2 (1+\lambda^2
K^2/d^2)$, $\omega_r=4\pi\sigma_{\perp}\epsilon$, $\Omega_r= 4\pi\sigma_{\|}$
and, finally,
$$\Pi=4\pi^2(\omega_p^2/i\omega)\sum_{l,m} \exp{[-i(l+m/2)qX
-imkZ/d]}.$$
Solution of equations (\ref{3}-\ref{4}) for the phase differences takes
the form
\begin{equation}
\varphi=\Pi(1+\lambda^2K^2/d^2)\delta(q-\omega/u)/D,
\label{fi}
\end{equation}
where $D$ is the determinant of equations (\ref{3}-\ref{4}). Zeros of $D$
yield the spectrum of the eigenmodes, {\em i. e.} of the plasmons. At high
frequencies and low temperatures, when frequencies of dielectric relaxation
$Re\,\omega_r$ è $Re\,\Omega_r$, are small enough, we get
$$D=[(\omega_p^2-\omega^2)(1+\lambda^2K^2/d^2)+\omega_p^2\lambda_c^2q^2
-i\omega[\omega_r(1+\lambda^2K^2/d^2) +\Omega_r(\omega_p^2-\omega^2
+\omega_p^2\lambda_c^2q^2)/\Omega_p^2].$$

Variation of the solution with frequency increasing can be illustrated by
the example of the component of the electric field $E_z=-i(\omega/2d)
\varphi$. At frequency $\omega=qu$ ($q=2\pi/X$) the slowly varying along $z$
part of the field $E_z$ created by a single vortex has the form
\begin{equation}
E_z=\frac{\pi u\omega_p^2 \exp{(-z/\Lambda})}{2d\sqrt{(\omega_p^2
-\omega^2-i\omega\omega_r)(\omega_p^2-\omega^2+\omega_p^2\lambda_c^2q^2
-i\omega\omega_r)}},\;\; \Lambda=\lambda\sqrt{\frac{\omega_p^2
-\omega^2-i\omega\omega_r}{\omega_p^2-\omega^2+\omega_p^2\lambda_c^2q^2-
i\omega\omega_r}}.
\label{Ez}
\end{equation}
The total field is described by the sum of the fields created by individual
vortices. One can see that at small frequencies the decaying length is of
the order of $\lambda$. When frequency becomes larger, than the plasma
frequency, the real part of the exponent in (\ref{Ez}) becomes small. This
means that in the range of plasma oscillations the decaying length for the
electric field is large and is determined by the damping of plasmons.

In order to find the relation between transport current $j_{tr}$ and the
velocity of the vortex lattice $u$, we multiply equation (\ref{1}) by
$(c^2/\lambda_c^2)\partial \varphi_n/\partial x$, and (\ref{2}) by
$4\partial P_n/\partial x$, then we add these equation, integrate them by
$x$ and sum up over all $n$. Then in the left-hand part of the resulting
equation we get an expression which is a full derivative over $x$, the
integral of which can be reduced to $-2d^2\sum_n H_y^2
\mid^{+\infty}_{-\infty}$. This expression is related to the force acting on
the vortex lattice and is proportional to the half sum of the magnetic
fields at the opposite sides of the sample ({\em i. e.} to the externally
applied magnetic field), and to the difference of the magnetic fields, which
by means of equation (\ref{M}) can be expressed in terms of the transport
current. The right-hand side we express by means of the Fourier
transformation using solutions of equations (\ref{3}-\ref{4}). Finally, we
get
\begin{equation}
16\pi^2dj_{tr}=\int dqdk
u\omega_p^4[\epsilon\omega_r(1+\lambda^2K^2/d^2)^2 +\Omega_r
\lambda^4K^2q^2/d^2)]/|D|^2\sum_{l,m}e^{i(l+m/2)X+imkZ/d}.
\label{ju}
\end{equation}
Integrating (\ref{ju}) we find a relation of the transport current to the
velocity of the vortex lattice, and, hence, to the average electric field
$\bar{E}=2\pi n_L u$ (where $n_L$ is the vortex density).

In the limit of small velocities, $u \ll (d/\lambda)c$, one can neglect the
effect of velocity $u$ on the shape of vortices. In this limit I-V curve is
described by the Ohm's law with the resistivity
\begin{equation}
\rho= \frac{1}{\sigma_\perp +\sigma_\|(\lambda/\lambda_c)^2}
\frac{2n_Ld\lambda_J}{\pi }.
\label{j1}
\end{equation}
where $\lambda_J=d\lambda_c/\lambda$. The last factor in (\ref{j1}) describes
the relative volume of the superconductor occupied by the non-linear region.
One can see an analogy to the Bardeen-Stephen law with the area of the
Abrikosov vortex core $\xi^2$ substituted by the area of the non-linear part
of the Josephson vortex $d \lambda_J$.

An analytic expression for the dependence of $u$ on $j_{tr}$ may be obtained
easily also for frequencies of the order of the plasma frequency for the
case of small plasmon damping
$$j_{tr}=j_c \frac{\sinh{b}}{\cosh{b} - \cos{a}},\;\; a=
\frac{Xc_z}{\lambda_c u}, \;b=\frac{\omega_r(\omega_p)}{2\omega_p}a.$$
The last factor in this equation describes the oscillations at the I-V curve
appearing as a result of interference of the plasmons created by different
vortices. In the limit of large velocities the I-V curve approaches the
Ohm's law again, but with resistivity $\rho=2\pi\sqrt{n_L d\lambda_j}
/\sigma(\omega_p)$. The shape of the I-V curves is sensitive to the
magnitude and to the anisotropy of the quasi-particle conductivity. An
increase of the conductivity results in an increase of the plasmon damping
and in a decrease of the peaks. The regime of very small damping is easily
achieved in the case of isotropic pairing when the order parameter has no
nodes and the quasi-particle density is exponentially small. In the case of
d-pairing the damping is higher and peaks at the I-V curves are smaller. In
Fig.1 we present the dependence $j_{tr}(u)$ calculated numerically under
assumptions that the real part of the conductivity is described by
expressions from \cite{lr} and is decreasing at frequencies above the
quasi-particle momentum scattering rate as $1/\omega^2$. Note that the
condition (\ref{d}) is difficult to satisfy at large currents, therefore the
shape of the I-V curve near the maxima in Fig.1 may need more detailed
calculations which take into account the bending of the vortices due to
their motion.

\begin{figure}
\vspace{85mm}
\caption{
Typical relation between normalized transport current and
vortex velocity in magnetic field  $H=20 H_{c1}$.}
\label{fig1}
\includegraphics{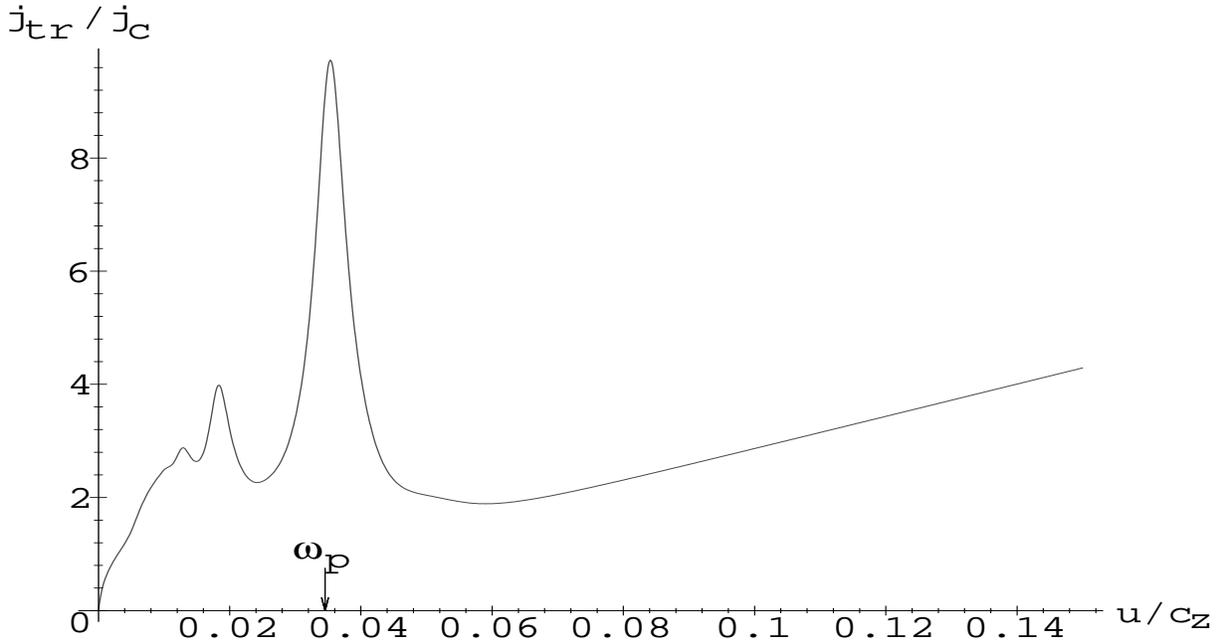}
\end{figure}

Thus, in the range of plasma oscillations the regions of negative
differential conductivity appear in which the uniform flow of the vortex
lattice is unstable. These regions are separated by the stability regions. At
large voltages the uniform motion is stable.

Excitation of plasma oscillations is related to an emission of
electromagnetic radiation. We calculate the energy flow traveling along $x$
axis at frequency $\omega_N = 2\pi u N/X$, where $N$ is the harmonic's
number. To do this we find the Pointing vector using expressions for
electric and magnetic fields calculated by means of (\ref{fi}).
$$S=\int^{\pi}_{-\pi} \frac{uc^2\omega_p^4(1+\lambda^2k^2/d^2)}{16\lambda
ZdX^2 |D(\omega=\omega_N)|^2} \sum_m e^{imZk/d} dk.$$

\begin{figure}
\vspace{85mm}
\caption{Typical dependence of energy flow radiated at the frequency
of the first harmonic, $H=20 H_{c1}$.}
\label{fig2}
\includegraphics{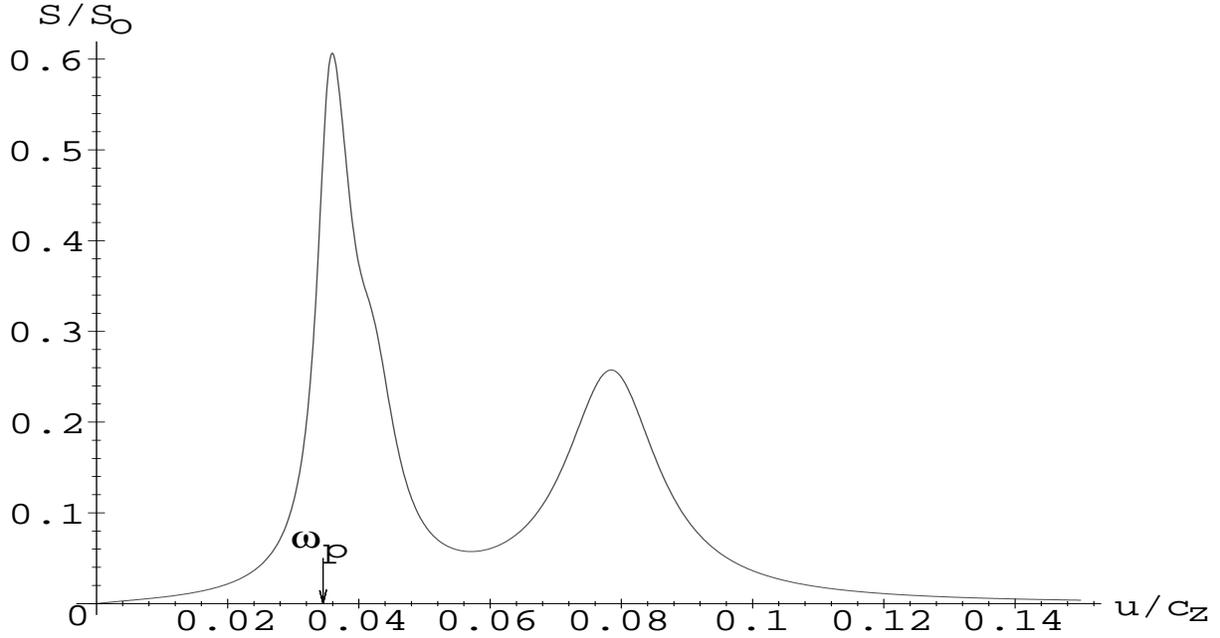}
\end{figure}

It is evident that the energy flow increases at resonance frequencies, when
the denominator $|D|$ is small. In the case of triangle lattice we find that
the energy of odd harmonics contains maxima in the region of plasma
frequency, while even harmonics has additional sharp peaks at the velocities
of the lattice close to the velocity of the light in the superconductor
$c_z$ playing a role of the Swihart velocity of the tunnel junction. The
latter peaks correspond to the Cherenkov radiation. The energy flow created
by the first harmonic, $S(u)$, is shown in Fig.2, where the energy flow is
presented in units of $S_0=\hbar^2c^2\omega_p/(16 e^2\lambda^2\lambda_c)$.
Using values typical for BSCCO, $\epsilon =25$, $\lambda =0,2$ $\mu$m,
$\lambda_c = 60$ $\mu$m, we find $S_0 \sim 10$ W/cm$^2$. The shape of the
curve and the peak's size strongly depend on the conductivity value and on
the anisotropy of the conductivity. The values of the lattice velocity
corresponding to the maximums at $S(u)$ do not necessarily correlate with
the peaks of the conductivity, thus, the peaks of the radiation in the range
of the plasma frequency can be observed in the regions of the stability of
the uniform vortex lattice flow.

The peak in the energy flow at $u >c_z$ is much higher than those near
the plasma frequency, it can be approximated by the expression
$$S=\frac{S_0 h^2}{4\pi^2\{[1+hN^2(1-u^2/c_z^2)]^2 +
hN^2\omega_r^2(u=c_z)\}},$$ where $h=4\pi^2\lambda\lambda_c n_L$.

Note that the frequencies of the radiation at the vortex motion with
velocities close to $c_z$ belong to the frequency range below the
superconducting gap value only if the magnetic field is small enough, {\it i.
e.} when the period of the vortex lattice $X$ is comparable to $\lambda_c$.
In larger fields the corresponding frequency approaches $\Delta$. However,
this must not affect the results considerably, because at such frequencies
the part of the response corresponding to damping is very small in pure,
$\tau T_c \gg 1$, superconductors.

Note also that the effects considered here are related to the range of higher
voltages and frequencies, than those typically used in experimental studies
of Josephson vortices, {\em e. g.} in \cite{U}, in which a non-Josephson
emission from the moving vortices was reported in BSCCO. As long as we
consider here the uniform motion of the lattice in the infinite crystal,
such a radiation in our treatment is absent.

This work was supported by Russian High-T$_c$ program under the grant No
96053.

\end{document}